\documentclass[runningheads]{llncs}
\usepackage[T1]{fontenc}
\usepackage{graphicx}
\usepackage{amsmath, amssymb}
\usepackage{mathtools}        
\usepackage{bm}
\usepackage{xcolor}
\usepackage{hyperref}
\usepackage{cleveref}
\usepackage{outlines}
\usepackage{booktabs}
\usepackage{array}
\usepackage{tabularx}
\usepackage{pifont}
\usepackage{makecell}

\newcommand{\D}{\mathsf{Def}}   
\newcommand{\A}{\mathsf{Adv}}   
\newcommand{\Lab}{\mathsf{Lab}}
\newcommand{\NOP}{\textsf{NOP}}

\newcommand{\EMB}{\textsf{EMBARGO}}
\newcommand{\PRE}{\textsf{PRE\_RELEASE}}
\newcommand{\PUB}{\textsf{PUBLIC}}

\newcommand{\ind}[1]{\mathbf{1}[#1]}
\newcommand{\val}{\operatorname{val}}

\DeclareMathOperator*{\argmax}{arg\,max}

\newcommand{\step}[1]{\ding{\the\numexpr191+#1\relax}} 
\newcommand{\stg}[1]{\mathsf{#1}}   
\newcommand{\act}[1]{\mathsf{#1}}   
\renewcommand{\NOP}{\act{nop}}

\definecolor{forestgreen}{RGB}{34, 139, 34}

\definecolor{mediumblue}{RGB}{0, 0, 205}

\begin{document}
\title{The Oracle's Gambit: A Game-Theoretic Framework for Responsible AI Release}
\titlerunning{The Oracle's Gambit}
%
\author{Christoph R. Landolt\inst{1}, Tobias Lorenz\inst{1}, Marta Kwiatkowska\inst{2}, Mario Fritz\inst{1}}
\authorrunning{C. R. Landolt et al.}
%
\institute{CISPA Helmholtz Center for Information Security 
\and
Department of Computer Science, University of Oxford
\email{christoph.landolt@cispa.de}\\
}
\maketitle              
\begin{abstract}
Responsible vulnerability disclosure can secure the defender’s head start by controlling when a vulnerability becomes public. However, this status quo is now challenged by increases in capability of AI models, which benefits both defenders and adversaries. When both sides draw their capability from the same AI model, the defender's head start depends on the lab's decision to release the model, and the question becomes not whether to release but how. Existing safety frameworks govern only the deploy-or-withhold threshold and leave the timing of release unmodeled. We cast this decision as a bilevel Stackelberg game in which a lab commits to a window that sets each side's capability over time in a downstream contest between defender and adversary. Defender welfare turns on the capability gap, not the shared level. Handing one model to both sides can trap the defender in a Red Queen's race, whereas a pre-release to the defender alone creates a protective gap, and the lab's optimal window balances this welfare gain against the opportunity cost of delaying release. For dual-use models, the lever is the sequencing of access, not the deployment threshold.

\end{abstract}

\vspace{-0.75em}
\section{Introduction}
\label{sec:introduction}
\vspace{-0.75em}
When a zero-day vulnerability surfaces, the adversary's path to damage is short: discover the flaw, weaponize it, strike. The defender's path to safety is longer: discover the same flaw, build a patch, test it, ship it, and wait for the fleet of systems to install it, a rollout that remains slow, costly, and bottlenecked by end-user adoption. This pipeline asymmetry means that, without some compensating defender advantage, the adversary finishes first.

Responsible disclosure is the established mechanism for providing that advantage to the defender. Arora et al. \cite{arora_optimal_2008} formalized the canonical model: a coordinator decides how long to withhold a vulnerability before public disclosure, while the software vendor decides when to ship a patch. The welfare-optimal policy is an interior deadline, long enough for the vendor to patch but short enough to incentivize action, and the 90-day window (with a possible 30-day extension) became the industry norm. The framework succeeds under three conditions: the flaw is not yet independently known to adversaries, patching is faster than independent rediscovery, and reverse-engineering a shipped patch into a working exploit takes long enough for the fleet to update. Together, these conditions mean that responsible disclosure creates a temporary information advantage for the defender, allowing them to ship a patch before exploitation is likely.

Each of these conditions is now under pressure from a common source: AI agents can discover the same vulnerability independently and cheaply across actors, dissolving the information advantage that coordinated disclosure was designed to create \cite{fang_llm_2024}. Reverse-engineering time from patch to exploit has collapsed from days or weeks to hours \cite{charrier_how_2024,epp_zero_2026}. And the time to weaponize a known flaw has dropped in tandem, as frontier models have demonstrated the ability to exploit real one-day vulnerabilities autonomously \cite{fang_llm_2024,wang2026exploitgymaiagentsturn,zhang_bountybench_2025,zhang_cybench_2024}.

These dynamics shift the disclosure decision upstream, to the laboratories that build frontier models and decide how to release them. OpenAI's Preparedness Framework \cite{openai_preparedness_2025}, Anthropic's Responsible Scaling Policy \cite{anthropic_responsible_2026}, and Google DeepMind's Frontier Safety Framework \cite{google_frontier_2026} each define capability thresholds at which deployment requires additional safeguards or must be withheld entirely. Yet these frameworks answer the binary question whether to deploy at all, not \emph{how} to responsibly release a model that clears the deployment threshold but still carries dangerous dual-use capability. The need for a more fine-grained approach to model release decisions beyond binary deployment thresholds is not hypothetical: evaluating Anthropic's Claude Mythos Preview, the UK AI Security Institute found it the first model to complete a simulated multi-step network intrusion end-to-end \cite{ai_security_institute_our_2026}, while Mozilla used the same model to identify and remediate 271~Firefox vulnerabilities before the model's public release \cite{holley_zero-days_2026}.

We recast this problem with a new leader and a new lever: the frontier laboratory, and the capability gap an AI model release induces. We model the decision as a bi-level three-player Stackelberg game in which the laboratory commits to a release policy that parameterizes a downstream zero-sum stochastic game between a defender and an adversary, solved to its unique value by backward induction. The laboratory then selects the policy that maximizes its payoff, trading downstream defender welfare against the opportunity cost of delaying release. The zero-day disclosure literature framed traditional disclosure timing as a game in which neither immediate disclosure nor secrecy is optimal \cite{arora_optimal_2008}; we move that decision upstream to the lab and make the lever a capability gap rather than an information gap.

Our central result is that \emph{symmetric capability scaling does not benefit the defender and can reduce its welfare, whereas the capability gap a pre-release opens raises it}. Releasing the same model to both sides at once equips them equally and gives neither a head start, yet the structural pipeline asymmetry favoring the adversary remains intact. A pre-release window restores the balance by opening a temporary capability gap: the defender operates at the frontier while the adversary remains a generation behind, rebuilding the time advantage that responsible disclosure once provided. Neither public release nor embargo opens the gap: public release equips both sides at the frontier, embargo holds both at the previous generation. Calibrating against real vulnerability data with capabilities elicited through an LLM-Delphi method \cite{lorenz_scalable_2026}, we find defender welfare and the lab's payoff are jointly maximized by a pre-release, robustly across the calibrated uncertainty range.

\medskip\noindent\textbf{Contributions.}
\begin{enumerate}
    \item \textbf{Frontier AI lab as a third-party leader over capabilities.} Unlike the usual defender-led security game, here a frontier AI laboratory leads by setting \emph{both} players' time-varying capabilities through its release policy, inducing a downstream zero-sum stochastic game between defender and adversary (\Cref{sec:model}).
    \item \textbf{The capability gap, not the level, drives welfare.} We show that defender welfare depends on the \emph{gap} between defender and adversary AI capability rather than on the shared level: handing both sides the same AI model traps the defender in a Red Queen's race, whereas a pre-release opens a protective gap that public release forgoes (\Cref{subsec:prerelease}).
    \item \textbf{Calibration to real frontier-model transitions.} We instantiate the game on successive AI model releases, with AI capabilities elicited by an LLM-Delphi panel, and show that a pre-release remains the lab's equilibrium choice across the elicited uncertainty, lowering the equilibrium attack frequency (\Cref{subsec:prerelease}).
\end{enumerate}

\vspace{-0.75em}
\section{Related Work}
\vspace{-0.75em}
Our framework spans several areas of prior work: the economics of vulnerability disclosure, game-theoretic models of cyber conflict, Stackelberg security games, the release and evaluation of frontier AI systems, and the quantitative modeling of AI risk through structured expert elicitation. We review each in turn.

\vspace{-0.75em}
\paragraph{Vulnerability Disclosure Economics.}
This literature frames disclosure as a timing problem controlled by whoever holds the patch. Rescorla \cite{rescorla_is_2005} finds no clear statistical evidence that finding and disclosing vulnerabilities substantially improves software quality, questioning whether the benefit outweighs the post-disclosure exposure. Arora et al. \cite{arora_optimal_2008} model a coordinator who sets a disclosure deadline and a vendor who chooses when to patch, showing that an interior deadline is welfare-optimal because it forces the vendor to patch sooner. Choi et al. \cite{choi_network_2010} let the vendor choose both its security investment and disclosure policy. Their model considers that an update only protects users who install it, while the disclosure itself enables adversaries to reverse-engineer the flaw. Closest to our setting, Canann \cite{canann_toward_2019} adds a profit-maximizing adversary and heterogeneous users, showing that disclosure improves welfare only when zero-days are cheap enough that the adversary attacks regardless and enough users still patch. This finding is our point of departure. As frontier AI drives Canann's search cost toward zero, the adversary attacks regardless, uncovering the flaw with or without disclosure. The protective timing that disclosure once provided dissolves, leaving as the defender's only head start the capability gap that opens when it holds the newest AI model while the adversary remains a generation behind.

\vspace{-0.75em}
\paragraph{Game-Theoretic Models of Cyber Conflict.} 
A parallel literature models the contest between adversary and defender directly, as a game over whether to attack, patch, or hold. Moore et al. \cite{moore_would_2010} weigh stockpiling a vulnerability against disclosing it, and Axelrod and Iliev \cite{axelrod_timing_2014} characterize the optimal moment to use a stockpiled exploit. Bao et al. \cite{bao_cyber_2021} model the vulnerability discovery, exploitation, and patching lifecycle as a two-player zero-sum partial-observation stochastic game, reduce it to a belief-state stochastic game solved by Shapley backward recursion, and validate it on the DARPA Cyber Grand Challenge. What this line of work shares is a closed contest in which capability is a fixed, exogenous parameter. AI breaks that closure. When adversary and defender draw their capability from the same released model, capability becomes endogenous, set by the lab's release decision. We add that lab as a third player, a leader whose release policy sets both sides' capabilities and turns this adversary-defender contest into its downstream subgame.

\vspace{-0.75em}
\paragraph{Stackelberg Security Games.} 
A common way to model the effect of defense measures and guardrails is the Stackelberg security game, in which a defender commits to a defensive strategy, and an adversary best responds. This paradigm has been developed and deployed at scale for allocating defensive resources \cite{tambe_security_2011,kamhoua_game_2021}, with the leader being the defender. Our leader is a third party, the releasing laboratory, which sets both players' capabilities through its release policy, then lets the defender and adversary compete to equilibrium.

\vspace{-0.75em}
\paragraph{Frontier-AI Release and Evaluation.}
Existing safety frameworks developed by frontier AI laboratories largely focus on whether to deploy an AI model at all, rather than how to release one that clears the deployment threshold but still carries dual-use capability \cite{openai_preparedness_2025,anthropic_responsible_2026,google_frontier_2026}. Prior work has explored several intermediate release mechanisms, ranging from the staged release of GPT-2 \cite{solaiman_release_2019} and Solaiman's later gradient-release methods \cite{solaiman_gradient_2023} to Shevlane's structured-access framework \cite{shevlane_structured_2022}. We build on this line of work by casting the pre-release channel as an economic decision: the lab trades the welfare gain of a protective gap against the opportunity cost of delaying public release, and optimizes the channel's duration rather than selecting it manually. The importance of release-channel design is underscored by growing evidence of frontier-model cyber capabilities, including demonstrations that leading models can exploit most tested one-day vulnerabilities \cite{fang_llm_2024} and strong performance on agentic attack-and-patch benchmarks involving real systems \cite{phuong_evaluating_2024,zhang_cybench_2024,zhang_bountybench_2025,wang_cybergym_2025,wang2026exploitgymaiagentsturn}. 

\vspace{-0.75em}
\paragraph{Quantitative AI Risk and Expert Elicitation.}
An open problem in frontier-AI release and evaluation is how to translate benchmark capabilities into quantitative risk. Murray et al. \cite{murray_mapping_2025} map benchmark scores to risk estimates through expert elicitation, and Barrett et al. \cite{barrett_toward_2025} decompose attacks into MITRE ATT\&CK steps and elicit AI uplift from human and LLM-simulated Delphi panels. Both use elicitation to \emph{measure} the risk a model poses, as input to a deploy-or-withhold decision. We instead use elicitation to choose \emph{when} to release rather than to score \emph{whether} to: the elicited capabilities set the per-round rates of defender and adversary on which our release game is solved, obtained from the LLM-Delphi procedure of Lorenz and Fritz \cite{lorenz_scalable_2026}, which adapts the Delphi method \cite{dalkey_experimental_1963} to language models and builds on evidence that LLM panels can approach human forecasting accuracy \cite{halawi_approaching_2024,schoenegger_wisdom_2024}.

\vspace{-0.75em}
\section{Model: Three-Player Stackelberg Security Game}
\label{sec:model}
\vspace{-0.75em}
Our framework models the inherently dual-use nature of frontier AI models in cybersecurity. These models can help software vendors detect new vulnerabilities and automate patch creation, while simultaneously enabling adversaries to automate the discovery and exploitation of zero-day vulnerabilities. Because both sides draw their capabilities from the same AI model, releasing the next-generation frontier model determines the capability profiles of \emph{both} players, eroding the classical timelines of responsible vulnerability disclosure and making the frontier AI lab a strategic actor in the vulnerability lifecycle. Our framework answers which release strategy a lab should follow to balance the system's welfare against its economic interests.

To analyze such a release, we formulate a bilevel, three-player Stackelberg game over a security subgame (\Cref{fig:Model_Overview}). The frontier AI lab $\Lab$ acts as the leader, committing to a release policy, while the defender $\D$ and adversary $\A$ are followers who interact in an induced stochastic game. The lab itself does not directly cause harm. Its committed policy instead parameterizes the interaction between $\D$ and $\A$ by setting both sides' access to the frontier model.

The game has two levels: the frontier AI lab, which controls a model's release, and the inner game between adversary and defender that the release induces. \Cref{subsec:overview} introduces this bilevel structure and the capabilities linking its two levels; \Cref{subsec:inner-sg} then details the inner game's dynamics and value, and \Cref{subsec:leader} the leader's policy and payoff.

\vspace{-0.75em}
\subsection{Game structure}
\label{subsec:overview}
\begin{figure}[t]
 \centering
 \includegraphics[width=\linewidth, trim={0.71cm 0.2cm 0.71cm 0.25cm}, clip]{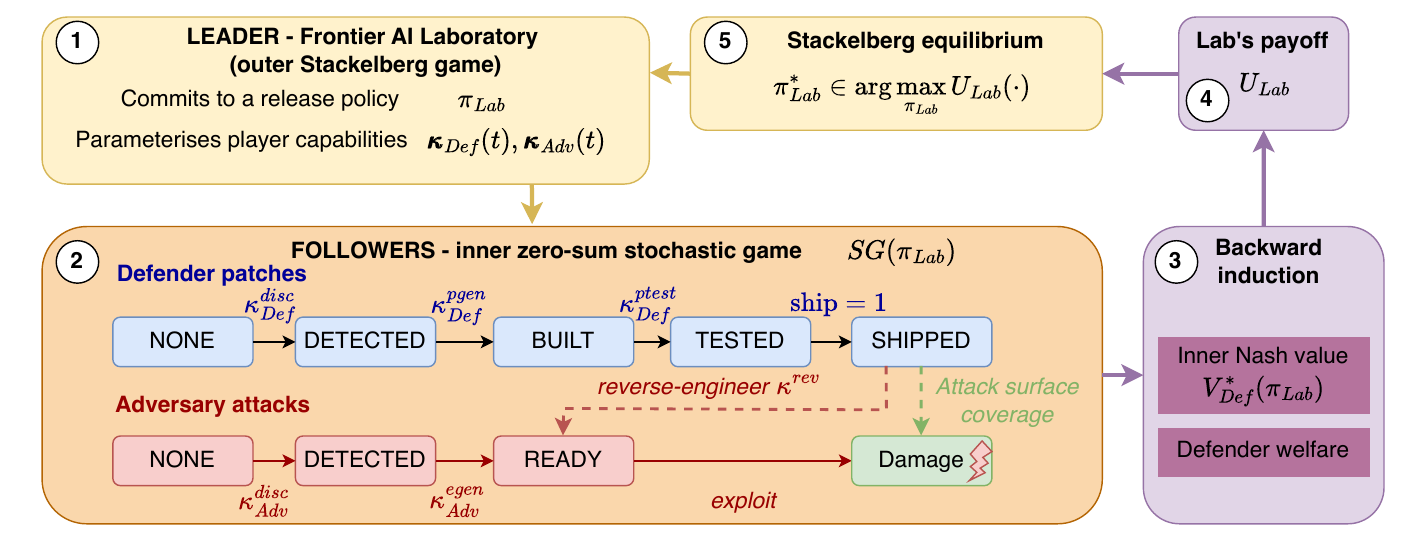}
 \caption{Overview of the bilevel game. The lab $\Lab$ (leader) commits to a release policy $\pi_{\Lab}$ that sets both players' capabilities (\step{1}), inducing an inner zero-sum stochastic game between the defender $\D$ and adversary $\A$ (\step{2}). Backward induction solves this game to its value (\step{3}), and the resulting defender welfare, net of the opportunity cost of delaying the release, forms the lab's payoff (\step{4}), whose maximization over policies gives the Stackelberg equilibrium (\step{5}).}
 \label{fig:Model_Overview}
\end{figure}

\vspace{-0.75em}
The game shown in \Cref{fig:Model_Overview} has the following structure. The release policy determines both players' capability schedules (\step{1}). Conditional on this policy, $\D$ and $\A$ play a fully observable, zero-sum stochastic game over the discovery, exploitation, and patching of $n$ vulnerabilities (\step{2}), whose unique value is obtained by backward induction (\step{3}). This value determines the lab's payoff (\step{4}), and maximizing it over admissible release policies yields the Stackelberg equilibrium (\step{5}).

We formalize this in the tuple
\begin{equation}
\mathcal{G} \;=\; \bigl\langle\, \{\Lab, \D, \A\},\ \mathcal{A}_{\Lab},\ \{\mathrm{SG}(\pi_{\Lab})\}_{\pi_{\Lab}\in\mathcal{A}_{\Lab}},\ U_{\Lab} \,\bigr\rangle,
\label{eq:game}
\end{equation}
in which $\Lab$ is the Stackelberg leader and commits to a release policy $\pi_{\Lab}\in\mathcal{A}_{\Lab}$ (\Cref{subsec:leader}) that induces an inner stochastic game $\mathrm{SG}(\pi_{\Lab})$ between the defender $\D$ and the adversary $\A$ (\Cref{subsec:inner-sg}), whose value determines the leader's payoff $U_{\Lab}$ (\Cref{subsec:leader}).

Conditional on the policy $\pi_{\Lab}$, the defender $\D$ and adversary $\A$ play the inner game
\begin{equation}
 \mathrm{SG}(\pi_{\Lab}) \;=\;
 \bigl\langle\,\{\D,\A\},\ \Theta,\
 \{\mathcal{A}_{\D},\mathcal{A}_{\A}\},\ \Phi,\ R\,\bigr\rangle ,
 \label{eq:sg}
\end{equation}
a simultaneous-move, fully observable, zero-sum, finite-horizon stochastic game over $n$ co-disclosed vulnerabilities and a horizon of $T$ rounds. 

We model the inner game as a race on each vulnerability: each player must pass through a sequence of steps to patch or to exploit it. The defender follows the patch-management lifecycle \cite{souppaya_guide_2022}, detecting the flaw, generating a patch, testing it, and shipping it, after which the patch protects the fleet only as users install it \cite{edgescan_vulnerability_2025}. The adversary follows the offensive stages of the cyber kill chain \cite{hutchins_intelligence-driven_2011}, detecting a vulnerability, generating an exploit, and exploiting it, with a second route that reverse-engineers a shipped patch by diffing\footnote{\emph{Diffing} compares the patched binary against the unpatched one to localize the change, which often pinpoints the very flaw the patch closes and shortens the path to a working exploit.} it back into an exploit \cite{brumley_automatic_2008}. The two pipelines are asymmetric in length: a real-world defender must balance security, functional correctness, and availability, so they test before shipping and then wait for adoption, whereas the adversary need only reach a working exploit. This structural asymmetry is the premise of our analysis.

\paragraph{Modeling assumptions.}
The structural properties of $\mathrm{SG}(\pi_{\Lab})$ are design choices to guarantee a unique game value via backward induction, with each acting as a conservative baseline. First, \emph{full observability} allows the defender to monitor the adversary's pipeline progress perfectly. This simplification overstates the defender's natural strength, ensuring that the protective effects we report derive strictly from the AI capability gap. Second, a \emph{zero-sum} payoff structure makes the adversary a pure damage-maximizer, and so provides an upper bound on harm, since a real adversary that also weighed its operational cost would attack less. Third, a \emph{finite horizon} isolates a single model-generation release window. We defer a broader discussion of these game-theoretic abstractions to Section \ref{sec:discussion_limitations}.

An AI model determines the capabilities of $\A$ and $\D$. Because AI assistance does not support each step of the two players equally, we represent their per-round capabilities by the capability vector
\begin{equation}
\bm{\kappa} =
\bigl(\kappa^{\mathrm{disc}},\kappa^{\mathrm{egen}},
\kappa^{\mathrm{rev}},\kappa^{\mathrm{pgen}},\kappa^{\mathrm{ptest}}\bigr)\in[0,1]^5,
\label{eq:cap-vector}
\end{equation}
These capability rates couple the two levels of $\mathcal{G}$. Conditioned on policy $\pi_{\Lab}$, the release determines each player’s per-action capabilities, thereby inducing the schedules $\bm{\kappa}_{\D}(t)$ and $\bm{\kappa}_{\A}(t)$ (\Cref{subsec:leader}). The transition dynamics of the inner game consume these schedules stage by stage (\Cref{subsec:inner-sg}).

\vspace{-0.75em}
\subsection{Inner game (Defender vs.\ Adversary)}
\label{subsec:inner-sg}
\vspace{-0.75em}
We now specify the inner game's components, unpacking \eqref{eq:sg} in turn: the state $\Theta$, the legal actions $\mathcal{A}_{\D}$ and $\mathcal{A}_{\A}$, the transition kernel $\Phi$, the reward $R$, and the value they induce.

\paragraph{State space.}
The state tracks each player's progress through its pipeline. For each vulnerability $v$ we track the two pipeline stages and the ship round $r^{(v)}$ ($\varnothing$ before ship). With a shared round counter $t$,
\begin{equation}
 \begin{aligned}
 \Theta \;=\;
 \underbrace{\{0,\dots,T\}}_{t}
 \;\times\;
 \prod_{v=1}^{n}
 \Bigl[\;
 &\underbrace{\{\stg{NONE},\stg{DETECTED},\stg{READY}\}}_{\sigma_{\A}^{(v)}}\\
 \times\;
 &\underbrace{\{\stg{NONE},\stg{DETECTED},\stg{BUILT},\stg{TESTED},\stg{SHIPPED}\}}_{\sigma_{\D}^{(v)}}\\
 \times\;
 &\underbrace{\bigl(\{0,\dots,T-1\}\cup\{\varnothing\}\bigr)}_{r^{(v)}}\;
 \Bigr].
 \end{aligned}
 \label{eq:state-space}
\end{equation}
The round counter advances by one each round, and the initial state $\theta_0$ has $t=0$ with every stage at $\stg{NONE}$ and every ship round at $\varnothing$. The joint state space grows exponentially in $n$, so we restrict to $n\le n_{\max}$.

\paragraph{Action sets and legality.}
Each action advances the acting player one pipeline stage per round on a single vulnerability, with the two exceptions: $\act{exploit}$ deals damage from $\stg{READY}$ without advancing, and $\act{reverse\_engineer}$ jumps straight to $\stg{READY}$ by diffing a shipped patch,
\begin{equation}
 \begin{aligned}
 \mathcal{A}_{\D} &= \{\NOP\}\cup\bigl(\{\act{detect},\act{create\_patch},\act{test\_patch},\act{ship\_patch}\}\times\{1,\dots,n\}\bigr),\\
 \mathcal{A}_{\A} &= \{\NOP\}\cup\bigl(\{\act{detect},\act{create\_exploit},\act{exploit},\act{reverse\_engineer}\}\times\{1,\dots,n\}\bigr).
 \end{aligned}
 \label{eq:action-sets}
\end{equation}
Each player takes one action per round. The no-operation action $\NOP$ is always legal, while every other action requires its prerequisite stage on $v$, and $\act{reverse\_engineer}$ additionally requires $r^{(v)}\neq\varnothing$. Because $\act{exploit}$ does not advance the stage, the adversary may exploit a $\stg{READY}$ vulnerability in every remaining round, with the damage decaying as coverage grows. Because the adversary acts once per round, $\act{create\_exploit}$ and $\act{reverse\_engineer}$ compete: once a patch ships, it weaponizes either its own discovery or the shipped patch.

\paragraph{Transitions.}
\label{subsec:transitions}
A release fixes how fast each side advances along its pipeline through the rates of \eqref{eq:cap-vector}, scheduled over time by the policy as $\bm{\kappa}_{\D}(t)$ and $\bm{\kappa}_{\A}(t)$ (\Cref{subsec:leader}). Each legal action then succeeds as an independent Bernoulli trial at the acting player's capability rate, a memoryless model of one attempt per round, with failure leaving the stage unchanged. The two pipelines advance independently, so $\Phi$ is their product, and for the legal pairings,
\begin{equation}
\begin{aligned}
 \Pr[\stg{DETECTED} \mid \stg{NONE},\, \act{detect}] &= \kappa^{\mathrm{disc}}_i(t),\\
 \Pr[\stg{READY} \mid \stg{DETECTED},\, \act{create\_exploit}] &= \kappa^{\mathrm{egen}}_{\A}(t),\\
 \Pr[\stg{READY} \mid \{\stg{NONE},\stg{DETECTED}\},\, \act{reverse\_engineer}] &= \kappa^{\mathrm{rev}}_{\A}(t),\\
 \Pr[\stg{BUILT} \mid \stg{DETECTED},\, \act{create\_patch}] &= \kappa^{\mathrm{pgen}}_{\D}(t),\\
 \Pr[\stg{TESTED} \mid \stg{BUILT},\, \act{test\_patch}] &= \kappa^{\mathrm{ptest}}_{\D}(t),\\
 \Pr[\stg{SHIPPED} \mid \stg{TESTED},\, \act{ship\_patch}] &= 1,
\end{aligned}
\label{eq:transitions}
\end{equation}
where $i \in \{\D, \A\}$ indexes the acting player and $\act{ship\_patch}$ at round $t$ deterministically records $r^{(v)}=t$. The independent route $\act{create\_exploit}$ is the adversary's only path before a patch ships; once the defender ships on $v$, the same event that begins patch coverage also exposes the patch to diffing, so the adversary may instead play $\act{reverse\_engineer}$ at rate $\kappa^{\mathrm{rev}}_{\A}$. Because the adversary takes one action per round, the patch route competes with the independent route rather than dominating it.

\paragraph{Rewards.}
The reward is nonzero only when the adversary exploits a $\stg{READY}$ vulnerability, where it equals the damage dealt, a loss to the defender and a gain to the adversary,
\begin{equation}
R(\theta, a_{\D}, a_{\A}) =
\begin{cases}
-\,d^{(v)}\,\bigl(1 - \mathrm{cov}^{(v)}(\theta)\bigr) & a_{\A}=\act{exploit}(v),\ \sigma^{(v)}_{\A}=\stg{READY},\\
0 & \text{otherwise.}
\end{cases}
\label{eq:reward}
\end{equation}
The per-round damage $d^{(v)}\ge 0$ is the cost of a fully exposed exploit, discounted by the unpatched share $1-\mathrm{cov}^{(v)}(\theta)$. Coverage is zero until ship, then grows at the adoption rate $\rho$,
\begin{equation} \mathrm{cov}^{(v)}(\theta) = \begin{cases} 0 & r^{(v)} = \varnothing,\\ h\bigl(t-r^{(v)}\bigr) & r^{(v)} \in \{0,\dots,T-1\}, \end{cases} \label{eq:coverage} \end{equation}
with $h(\Delta t)=\min\bigl(1,\,\rho\max(0,\Delta t)\bigr)$. Because $R$ gates on $\sigma^{(v)}_{\A}=\stg{READY}$, attacks turn on exploit-readiness rather than disclosure in every regime, and because $R$ is independent of $a_{\D}$, the defender acts only through the transitions that drive a patch to ship. AI inference costs enter as a per-action cost $c_{\D}\ge 0$ in the defender's external reward,
\begin{equation}
 R^{\mathrm{ext}}_{\D} = R - c_{\D}\,\ind{a_{\D}\neq\NOP},
 \label{eq:reward-ext}
\end{equation}
which supplies the welfare term of $U_{\Lab}$ (\Cref{subsec:leader}). Because the inner game is solved on $R$ alone, this cost leaves the inner equilibrium unchanged.

\paragraph{The inner value.}
Being fully observable, zero-sum, and finite-horizon, the game admits a unique value, which we compute by backward induction over the horizon, the finite-horizon specialization of the stochastic game of Shapley \cite{shapley_stochastic_1953}. Writing $V(\theta)$ for the defender's value at state $\theta$, the recursion forms the stage matrix and takes its minimax value,
\begin{equation}
 M(\theta)_{a_{\D},\, a_{\A}}=R(\theta, a_{\D}, a_{\A})+\sum_{\theta'} \Phi(\theta, a_{\D}, a_{\A})[\theta']\, V(\theta'),
 \label{eq:stage-matrix}
\end{equation}
\begin{equation}
 V(\theta)=\val\bigl[M(\theta)\bigr]=
 \max_{x \in \Delta(\mathcal{A}_{\D})}\;
 \min_{y \in \Delta(\mathcal{A}_{\A})}\;
 x^{\!\top} M(\theta)\, y,
 \label{eq:shapley}
\end{equation}
with $V(\theta)=0$ at $t=T$. We write $V^*_{\D}(\pi_{\Lab}) := V(\theta_0;\pi_{\Lab})$ for the value at the initial state. The value is unique, though the optimal strategies need not be, which we revisit when defining welfare (\Cref{subsec:leader}).

\vspace{-0.75em}
\subsection{The leader: release policy, payoff, and equilibrium}
\label{subsec:leader}
\vspace{-0.75em}
\Cref{subsec:inner-sg} solves the inner game for a specified capability schedule. The leader's task is to condition that schedule through the policy $\pi_{\Lab}$, and to weigh the welfare it produces against the cost of delay. The rest of this subsection follows that order: the lever, the welfare it yields, the payoff, and the equilibrium.

\paragraph{Release policy.}
The leader's ($\Lab$) action is to choose when a new frontier model reaches $\D$ and $\A$ (\step{1}). It commits to a window length $W$ that determines how long the new model is withheld from the adversary,
\begin{equation}
\pi_{\Lab} = W \in \{0,1,\ldots,T\} \cup \{\EMB\}.
\label{eq:lab-action}
\end{equation}
Three regimes follow. A public release ($\PUB$), $W=0$, grants both sides access simultaneously. A pre-release ($\PRE$), $W\in\{1,\ldots,T\}$, grants the defender immediate access and the adversary access only after $W$ rounds, at which point the model becomes public. An embargo ($\EMB$) withholds the model from both sides for the entire horizon.

\paragraph{Induced capabilities.}
The release exposes two levels of the capability vector \eqref{eq:cap-vector}, the previous generation $\bm{\kappa}^{\mathrm{prev}}$ and the new one $\bm{\kappa}^{\mathrm{new}}$, with $\bm{\kappa}^{\mathrm{prev}} \le \bm{\kappa}^{\mathrm{new}}$ componentwise; a release moves a player from $\bm{\kappa}^{\mathrm{prev}}$ to $\bm{\kappa}^{\mathrm{new}}$ when its access opens. The window therefore sets the schedules $\bm{\kappa}_{\D}(t)$ and $\bm{\kappa}_{\A}(t)$ that the inner game consumes,
\begin{equation}
 \begin{aligned}
 \bm{\kappa}_{\D}(t) &=
 \begin{cases}
 \bm{\kappa}^{\mathrm{prev}} & \pi_{\Lab} = \EMB,\\
 \bm{\kappa}^{\mathrm{new}}  & \text{otherwise,}
 \end{cases}\\[2pt]
 \bm{\kappa}_{\A}(t) &=
 \begin{cases}
 \bm{\kappa}^{\mathrm{prev}} & \pi_{\Lab} = \EMB \ \text{or}\ t < W,\\
 \bm{\kappa}^{\mathrm{new}}  & \text{otherwise.}
 \end{cases}
 \end{aligned}
 \label{eq:release-event}
\end{equation}
The induced gap $\Delta\bm{\kappa}(t)=\bm{\kappa}_{\D}(t)-\bm{\kappa}_{\A}(t)$ is nonzero only inside a pre-release window ($t<W$), where the defender already operates at the frontier $\bm{\kappa}^{\mathrm{new}}$ while the adversary remains at $\bm{\kappa}^{\mathrm{prev}}$; public release and embargo both leave it at zero. \Cref{sec:calibration} shows that it is this gap, not the shared capability level, that moves defender welfare. 

\paragraph{Defender welfare.}
Under the inner equilibrium the defender realizes welfare
\begin{equation}
 u_{\D}(\pi_{\Lab})
 = \mathbb{E}\Bigl[\textstyle\sum_{t=0}^{T-1} R^{\mathrm{ext}}_{\D}\Bigr]
 = V^*_{\D}(\pi_{\Lab}) - c_{\D}\,\mathbb{E}\bigl[\,\bigl|\{t : a_{\D}\neq\NOP\}\bigr|\,\bigr],
 \label{eq:welfare}
\end{equation}
that is, the inner value $V^*_{\D}(\pi_{\Lab})$ minus a per-round action cost $c_{\D}$ for each time step in which the defender acts. Because the inner game is solved with respect to $R$ alone, the value $V^*_{\D}$ is unique, but the optimal defender strategy need not be (\Cref{subsec:inner-sg}). As a result, the induced action count, and hence $u_{\D}$, is not determined by $V^*_{\D}$ alone. In particular, value-optimal strategies that differ only on vulnerabilities never brought to $\stg{READY}$ by the adversary are payoff-equivalent under $R$ but differ under $R^{\mathrm{ext}}_{\D}$. To obtain a unique welfare measure, we select among value-optimal strategies the one with minimal expected action count. This is equivalent to taking the limit $c_{\D}\to 0^{+}$, which yields a unique $u_{\D}(\pi_{\Lab})$.

\paragraph{The lab's payoff.}
The lab weighs the defender welfare $u_{\D}(\pi_{\Lab})$ against the opportunity cost of withholding the model,
\begin{equation}
 U_{\Lab}(\pi_{\Lab})
 =
 \underbrace{u_{\D}(\pi_{\Lab})}_{\text{welfare}}
 -
 \underbrace{c_{\mathrm{delay}}\, \ell(\pi_{\Lab})}_{\text{opportunity cost}},
 \label{eq:UL}
\end{equation}
where $c_{\mathrm{delay}}\ge 0$ is the per-round cost of holding the model private and $\ell(\pi_{\Lab})$ is the number of rounds it is withheld from the adversary: $\ell=0$ under public release, $\ell=W$ under a pre-release of length $W$, and $\ell=T$ under embargo. The pre-release window thus trades the welfare gain of a wider protective gap against a delay cost that grows linearly in its length.

\paragraph{Stackelberg equilibrium.}
The lab commits first and anticipates the followers' equilibrium response, so the Stackelberg equilibrium is the policy that maximizes its payoff,
\begin{equation}
 \pi_{\Lab}^{*} \in
 \argmax_{\pi_{\Lab} \in \mathcal{A}_{\Lab}}\;
 U_{\Lab}(\pi_{\Lab}).
 \label{eq:stackelberg}
\end{equation}
Because $\mathcal{A}_{\Lab}=\{0,1,\dots,T\}\cup\{\EMB\}$ is finite, we solve \eqref{eq:stackelberg} by enumeration: for each candidate we solve the induced inner game to its value $V^*_{\D}(\pi_{\Lab})$ by backward induction (\Cref{subsec:inner-sg}), read off the welfare $u_{\D}(\pi_{\Lab})$, and select the policy of greatest payoff $U_{\Lab}(\pi_{\Lab})$.

\vspace{-0.75em}
\section{Game Calibration}
\label{sec:calibration}
\vspace{-0.75em}
This section grounds the model in real-world data and exercises it on real frontier-model releases. We sort every parameter into three classes: the fixed environment, the cost levers, and the AI capabilities---and ground each in the appropriate source. The AI capabilities are not directly observable as per-round rates, so an LLM-Delphi panel elicits them from benchmark evidence. A sensitivity analysis then identifies which components the game responds to (\Cref{subsec:parameterization}). We show empirically across the elicited model releases that releasing publicly leads to a Red Queen's race, whereas a temporary capability gap improves welfare: protection depends on the gap, not the shared level (\Cref{subsec:prerelease}).

\vspace{-0.75em}
\subsection{Parameterization and Game Dynamics}
\label{subsec:parameterization}
\vspace{-0.75em}
The game's parameters fall into three classes (\Cref{tab:params}). The \emph{Class A} environment is AI-independent: we calibrate the adoption rate to published remediation data and fix the horizon, surface size, and damage numeraire by modeling choice. The \emph{Class B} cost levers, $c_{\mathrm{delay}}$ and $c_{\D}$, are swept to locate the band where the leader's commitment is non-trivial and to anchor the operating point for the release decision. The \emph{Class C} capabilities, the five components of $\bm{\kappa}$, are not observable as transition probabilities and are therefore elicited from an LLM-Delphi panel \cite{lorenz_scalable_2026} that converts benchmark evidence into the per-round rates the model requires.

\begin{table}[t]
\centering
\caption{Parameter inventory by class: \textbf{A} environment, \textbf{B} swept cost levers, \textbf{C} capabilities (elicited, release-dependent). \emph{Defined} is the domain the implementation admits and \emph{Calibrated} the value or range used in the runs.}
\label{tab:params}
\footnotesize
\setlength{\tabcolsep}{4pt}
\begin{tabularx}{\linewidth}{@{}l >{\raggedright\arraybackslash}X c c l@{}}
\toprule
Symbol & Description & Defined & Calibrated & Source \\
\midrule
\multicolumn{5}{@{}l}{\textit{Class A --- environment}}\\[2pt]
$\rho$       & Post-ship fleet coverage rate & $(0,1]$ & $0.11$ & Edgescan \cite{edgescan_vulnerability_2025} \\
$d^{(v)}$    & Per-round damage (numeraire) & $\ge 0$ & $1.0$ & Unitcosts \\
$T$          & Per-window horizon (rounds) & $\ge 1$ & $48$ & Model Choice \\
$n$          & Co-disclosed vulnerabilities & $\mathbb{N}^+$ & $\{1,2,3\}$ & Model Choice \\
\addlinespace[3pt]
\multicolumn{5}{@{}l}{\textit{Class B --- swept cost levers}}\\[2pt]
$c_{\mathrm{delay}}$ & Per-round hold cost & $\ge 0$ & $[0,0.5]$ & Parameter Sweep \\
$c_{\D}$     & Defender per-action cost & $\ge 0$ & $[0,1.3]$ & Parameter Sweep \\
\addlinespace[3pt]
\multicolumn{5}{@{}l}{\textit{Class C --- capabilities (elicited, release-dependent)}}\\[2pt]
$\kappa^{\mathrm{disc}}$  & $\act{detect}$ (both sides) & $[0,1]$ & elicited & LLM-Delphi \\
$\kappa^{\mathrm{egen}}$  & $\act{create\_exploit}$  & $[0,1]$ & elicited & LLM-Delphi \\
$\kappa^{\mathrm{rev}}$   & $\act{reverse\_engineer}$ (patch diff) & $[0,1]$ & elicited & LLM-Delphi \\
$\kappa^{\mathrm{pgen}}$  & $\act{create\_patch}$ & $[0,1]$ & elicited & LLM-Delphi \\
$\kappa^{\mathrm{ptest}}$ & $\act{test\_patch}$ & $[0,1]$ & elicited & LLM-Delphi \\
\bottomrule
\end{tabularx}
\end{table}

\vspace{-0.75em}
\paragraph{Environment (Class A).}
The horizon $T$ spans a single release window, and the surface holds $n$ different vulnerabilities, both kept small ($n\le3$) for tractability. The per-round damage $d^{(v)}$ is the numeraire. The adoption rate $\rho$ governs how fast a shipped patch reaches the fleet through $h(\Delta t)=\min(1,\rho\Delta t)$, so coverage is full after $1/\rho$ rounds. Because $\rho$ is exogenous to both players, it is the one Class A parameter we calibrate from real-world data rather than set by design: a round is seven days, and the software-sector mean time to remediate, about $63$ days \cite{edgescan_vulnerability_2025}, gives $63/7=9$ rounds, which we match to $1/\rho$ to yield $\rho\approx0.11$.

\begin{figure}[t]
  \centering
  \includegraphics[width=\linewidth, trim={0.15cm 0.15cm 0.2cm 0.15cm}, clip]{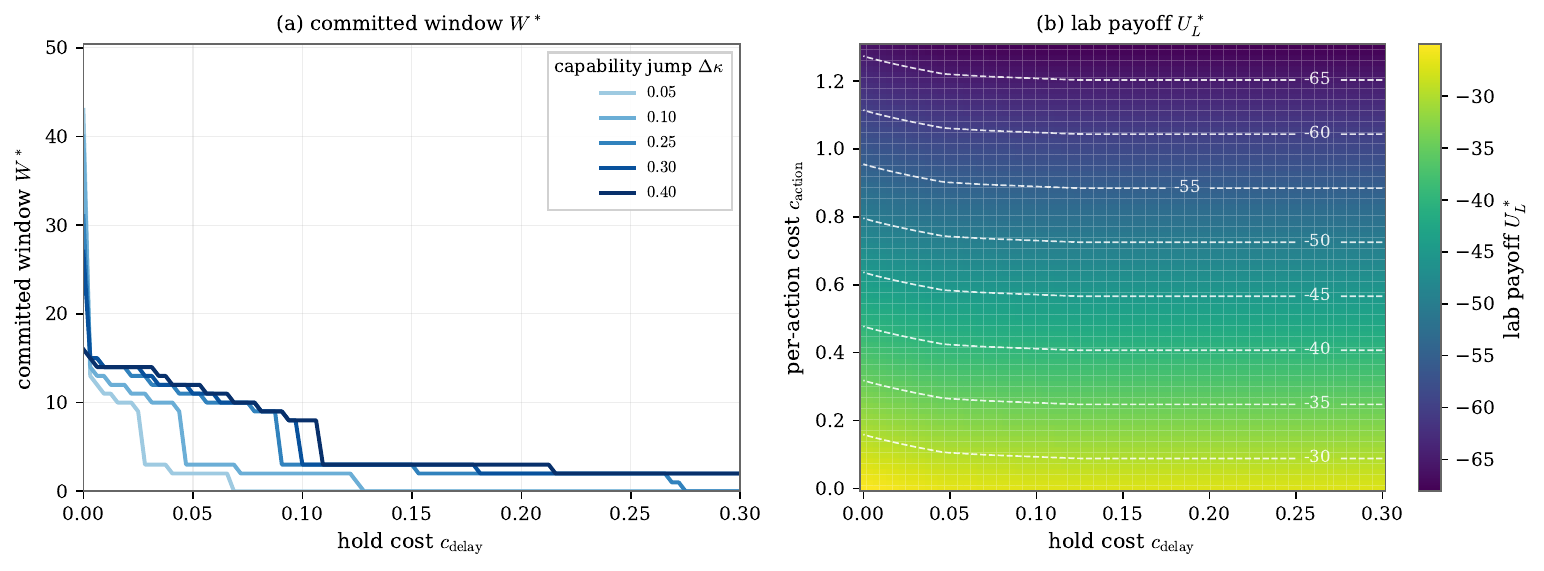}
\caption{Committed window under the two cost levers. (a) Optimal window $W^\star$ versus delay cost $c_{\mathrm{delay}}$ for several capability jumps $\Delta\bm{\kappa}$. As delay costs increase, $W^\star$ decreases in discrete steps, while larger capability jumps widen the optimal window. (b) Optimized lab payoff $U_{\Lab}^\star$ as a function of $c_{\mathrm{delay}}$ and the per-action cost $c_{\D}$. Although $c_{\D}$ does not affect the inner equilibrium, it enters the outer optimization objective and therefore shifts the optimal committed window. Base point: $\Delta\bm{\kappa}=0.10$, $n=2$, and $c_{\mathrm{delay}}=c_{\D}=0.05$.} 
\label{fig:commitment}
\end{figure}

\vspace{-0.75em}
\paragraph{Cost levers (Class B).}
Because cost levers are lab- and release-specific and are not directly observable, we sweep them to identify the band in which the leader's commitment decision is non-trivial (\Cref{fig:commitment}). We use a synthetic AI capability instance with $\bm{\kappa}^{\mathrm{prev}}=(0.64,\dots,0.64)$, $\bm{\kappa}^{\mathrm{new}}=(0.74,\dots,0.74)$, and hence a uniform capability increase of $\Delta\bm{\kappa}=0.10$, with $n=2$ vulnerabilities. The committed window $W^\star$ is interior only within a band of the delay cost $c_{\mathrm{delay}}$, and the per-action cost $c_{\D}$ shifts it within that band. We set $c_{\mathrm{delay}}=c_{\D}=0.05$, with $c_{\mathrm{delay}}$ at an interior point of the band.

\begin{figure}[t]
  \centering
  \includegraphics[width=\linewidth, trim={0.15cm 0.15cm 0.2cm 0.15cm},clip]{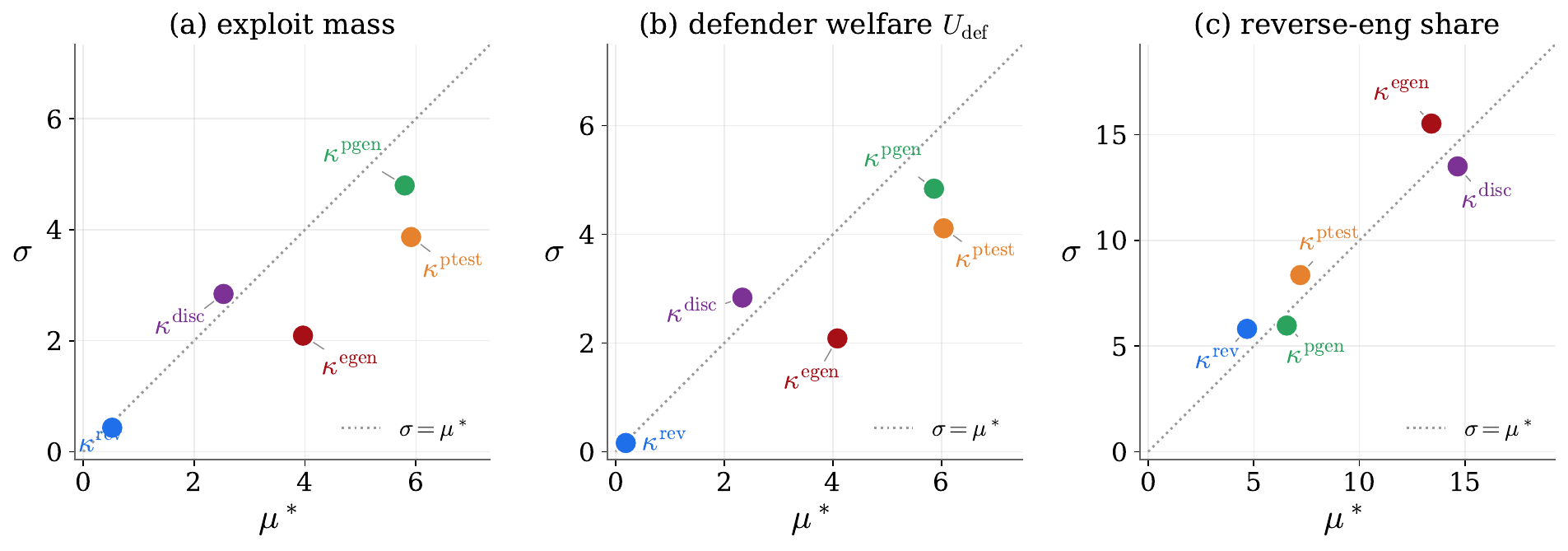}
\caption{Global capability sensitivity (Morris elementary effects). For each component of $\bm{\kappa}$, the mean absolute effect $\mu^\ast$ against its spread $\sigma$, on (a) attack success, (b) defender welfare, and (c) the patch-diffing share. Defensive rates $\kappa^{\mathrm{ptest}}$, $\kappa^{\mathrm{pgen}}$ lead on attack success and welfare; offensive $\kappa^{\mathrm{disc}}$, $\kappa^{\mathrm{egen}}$ on the patch-diffing share; $\kappa^{\mathrm{rev}}$ weakest throughout. Spreads $\sigma\!\approx\!\mu^\ast$ indicate interaction-heavy effects.}
  \label{fig:cap-morris}
\end{figure}

\vspace{-0.75em}
\paragraph{Capabilities (Class C).}
The AI capabilities are the last class to fix and the one that drives the inner game's dynamics, yet the five components of $\bm{\kappa}$ are not directly observable, so we elicit them rather than measure them. We precede the elicitation with a screening of which components the game responds to, applying the Morris method of elementary effects across three outcomes: attack success, defender welfare, and reliance on the patch-diffing route, the adversary's alternative to independent weaponization once a patch ships (\Cref{fig:cap-morris}). Patch testing $\kappa^{\mathrm{ptest}}$ and patch generation $\kappa^{\mathrm{pgen}}$ are most influential on attack success and defender welfare: the defender's longer pipeline is the binding constraint. Exploit generation $\kappa^{\mathrm{egen}}$ leads the offensive components and governs the patch-diffing share alongside discovery $\kappa^{\mathrm{disc}}$; reverse-engineering $\kappa^{\mathrm{rev}}$ is weakest throughout. The defensive rates and $\kappa^{\mathrm{egen}}$ thus drive the results, while $\kappa^{\mathrm{rev}}$ matters least.

\vspace{-0.75em}
\paragraph{Eliciting Class C.}
Our elicitation follows the Scalable Delphi method of Lorenz and Fritz \cite{lorenz_scalable_2026}, which has three parts: the target variables to estimate (here the five components of $\bm{\kappa}$), a board of LLM experts, and a shared evidence base. The board consists of five security-expert personas: a defensive specialist, a malware reverse engineer, an AI/ML security researcher, a threat-intelligence analyst, and a compliance officer. Each runs on GPT-5.1, so no assessed model judges itself. The evidence is a shared corpus of offensive benchmarks (CyberGym \cite{wang_cybergym_2025}, ExploitGym \cite{wang2026exploitgymaiagentsturn}, BountyBench \cite{zhang_bountybench_2025}, AISI \cite{ai_security_institute_our_2026}, Cybench \cite{zhang_cybench_2024}), defensive benchmarks (SWE-bench \cite{jimenez2024swebench}, FeedbackEval \cite{dai2025feedbackeval}, SWT-bench \cite{NEURIPS2024_94f093b4}), and the Zero Day Clock \cite{epp_zero_2026} for patch-to-exploit latency, together with each rated model's own system card. From this evidence, each panelist estimates the Bernoulli transition rates for the five successive frontier models (GPT-4o, OpenAI o4-mini, Claude Opus 4.5, Claude Opus 4.6, Claude Mythos Preview), mapping the benchmarks' end-to-end pass rates to the per-round rate of a single pipeline action (\Cref{subsec:transitions}), with a confidence and a rationale.

\begin{table}[t]
\centering
\caption{Elicited per-round capabilities (panel mean $\pm$ std).
Each cell is the Bernoulli transition rate for one pipeline action
per round.}
\label{tab:delphi-capabilities}
\footnotesize
\setlength{\tabcolsep}{4pt}
\begin{tabular}{lccccc}
\toprule
Model & $\kappa^{\mathrm{disc}}$ & $\kappa^{\mathrm{egen}}$ & $\kappa^{\mathrm{rev}}$ & $\kappa^{\mathrm{pgen}}$ & $\kappa^{\mathrm{ptest}}$ \\
\midrule
GPT-4o                & $0.07 \pm 0.03$ & $0.04 \pm 0.02$ & $0.06 \pm 0.03$ & $0.09 \pm 0.05$ & $0.12 \pm 0.06$ \\
OpenAI o4-mini        & $0.12 \pm 0.03$ & $0.07 \pm 0.03$ & $0.10 \pm 0.03$ & $0.15 \pm 0.05$ & $0.20 \pm 0.06$ \\
Claude Opus 4.5       & $0.17 \pm 0.04$ & $0.12 \pm 0.03$ & $0.15 \pm 0.03$ & $0.22 \pm 0.06$ & $0.28 \pm 0.06$ \\
Claude Opus 4.6       & $0.22 \pm 0.04$ & $0.14 \pm 0.04$ & $0.17 \pm 0.02$ & $0.24 \pm 0.06$ & $0.29 \pm 0.06$ \\
Mythos Preview & $0.31 \pm 0.06$ & $0.26 \pm 0.08$ & $0.29 \pm 0.06$ & $0.31 \pm 0.06$ & $0.37 \pm 0.07$ \\
\bottomrule
\end{tabular}
\end{table}

\begin{figure}[t]
  \centering
  \includegraphics[width=\linewidth, trim={0.15cm 0.15cm 0.2cm 0.15cm}, clip]{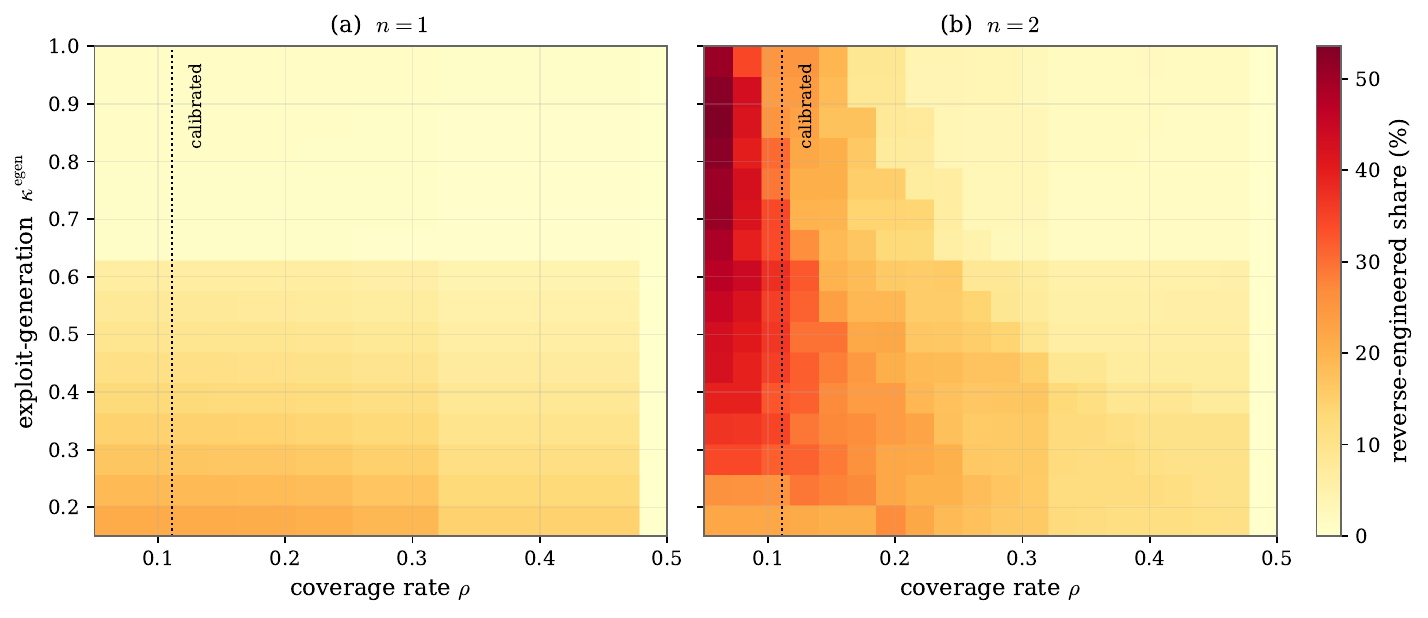}
\caption{Patch-diffing share over $\rho$ and $\kappa^{\mathrm{egen}}$, for (a) $n=1$ and (b) $n=2$. Governed by coverage (peaking at low $\rho$, gone by $\rho\gtrsim0.45$); its $\kappa^{\mathrm{egen}}$-dependence flips with the surface, and the peak rises from $21\%$ to $54\%$. Dotted line: calibrated $\rho$.}
  \label{fig:reveng-regime}
\end{figure}

\Cref{tab:delphi-capabilities} reports the consensus vectors (panel mean $\pm$ inter-expert standard deviation). Defensive actions succeed at higher per-round rates than offensive ones, yet the defender's longer pipeline absorbs this advantage. All rates rise monotonically across the model ladder, but the latest release (Opus 4.6 to Mythos Preview) concentrates its gains on the offensive side, most consequentially on $\kappa^{\mathrm{egen}}$, the most influential of the offensive components. The welfare consequences of this offense-weighted release are taken up in \Cref{subsec:prerelease}.
\vspace{-0.75em}
\paragraph{Game dynamics.}
Once a patch ships, the adversary chooses between weaponizing its own find at $\kappa^{\mathrm{egen}}$ and diffing the shipped patch at $\kappa^{\mathrm{rev}}$ (\Cref{subsec:transitions}); the diff route is gated on a shipped patch but bypasses discovery. Its share (\Cref{fig:reveng-regime}) is governed by coverage, peaking at low $\rho$ and vanishing once fast coverage closes the post-ship window ($\rho\gtrsim0.45$). Its dependence on $\kappa^{\mathrm{egen}}$ flips with the surface: confined to low $\kappa^{\mathrm{egen}}$ at $n=1$, where fast self-weaponization pre-empts any patch, but spanning the full range at $n=2$, where one action per round lets a patch shipped on one vulnerability be diffed while the adversary works another. Extra targets thus feed the diff route rather than crowding it out, and its peak rises from $21\%$ at $n=1$ to $54\%$ at $n=2$. At the calibrated $\rho\approx0.11$, with $\kappa^{\mathrm{rev}}\!\gtrsim\!\kappa^{\mathrm{egen}}$ across the elicited ladder (\Cref{tab:delphi-capabilities}), the route is material and the adversary genuinely two-route, with the two routes near-substitutes there, so outcomes stay insensitive to $\kappa^{\mathrm{rev}}$ itself (\Cref{fig:cap-morris}), bearing out the classical worry that a shipped patch is itself an exploit blueprint \cite{flake_structural_2004}.

\vspace{-0.75em}
\subsection{Case Study: Releasing a Frontier Model}
\label{subsec:prerelease}
\begin{figure}[t]
  \centering
  \includegraphics[width=\linewidth]{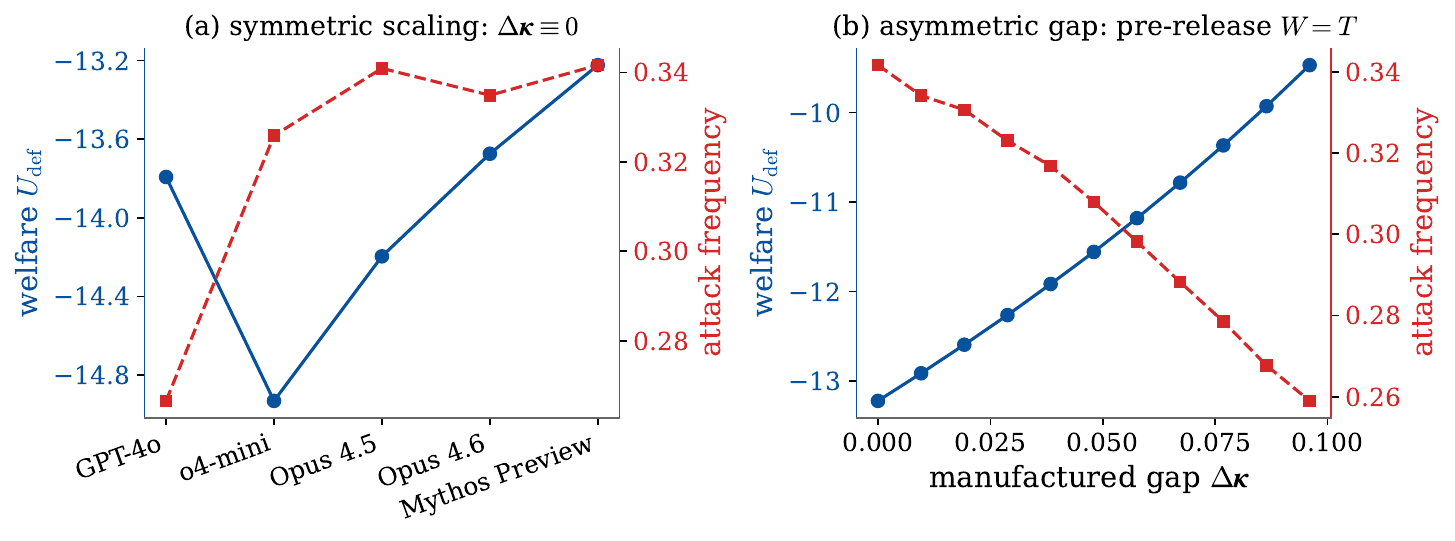}
\caption{Welfare turns on the capability \emph{gap}, not the \emph{level}. (a) Symmetric scaling ($\Delta\bm{\kappa}\equiv0$): raising the shared level yields no net welfare gain up the ladder, while attacks rise. (b) Asymmetric gap: defender fixed at the Mythos frontier, adversary swept back toward Opus 4.6.}
  \label{fig:scaling}
\end{figure}

\vspace{-0.75em}
Based on the calibrated game (\Cref{subsec:parameterization}), we analyze the lab's release decision across the elicited model ladder (\Cref{tab:delphi-capabilities}), solving every transition. We first compare welfare under public release against the lab's optimal strategy, showing that welfare turns on the capability gap, not the level. We then test that strategy's robustness to the epistemic uncertainty of the LLM-Delphi panel by Monte Carlo simulation. Finally, we examine the latest Opus 4.6 $\to$ Mythos Preview release in detail, relating the game's predicted strategy to Anthropic's real-world release decision.
\vspace{-0.75em}
\paragraph{Welfare turns on the capability gap, not the level.}
We compare five elicited frontier models, GPT-4o, OpenAI o4-mini, Claude Opus 4.5, Claude Opus 4.6, and Claude Mythos Preview, under two release regimes (\Cref{fig:scaling}a). A public release makes the newest AI model available to both defender and adversary, while a pre-release provides it to the defender alone, leaving the adversary one generation behind. Under symmetric scaling ($\Delta\bm{\kappa}\equiv0$, with the parameters from \Cref{subsec:parameterization}), attack frequency rises from $0.27$ to $0.34$ as the shared model advances from GPT-4o to Mythos, while defender welfare moves little on net, from $-13.79$ to $-13.22$. Scaling both sides equally intensifies attacks without improving defender welfare, a Red Queen's race driven by the pipeline-length asymmetry.

For a single vulnerability, the expected time to traverse a pipeline is $\sum_i 1/\kappa_i$, with one geometric wait for each Bernoulli stage, and the adversary reaches $\stg{READY}$ sooner than the defender reaches $\stg{SHIPPED}$. Increasing shared capability allows both players to speed up their pipeline transitions, but the same speed-up moves the adversary into a damaging state earlier. Because shipping does not immediately eliminate exposure, coverage grows at the capability-independent rate $\rho$, so an earlier-ready adversary accumulates damage over more rounds before the fleet is patched.

A pre-release instead opens a capability gap (\Cref{fig:scaling}b), with the defender operating at $\bm{\kappa}^{\mathrm{new}}$ while the adversary remains at $\bm{\kappa}^{\mathrm{prev}}$, so the defender can ship and begin accumulating coverage before the adversary reaches $\stg{READY}$. In our calibration, this gap raises defender welfare from $-13.22$ to $-9.47$ and lowers attack frequency from $0.34$ to $0.26$. The welfare gain thus comes from relative capability, not from a higher common level.

\begin{figure}[t]
  \centering
  \includegraphics[width=\linewidth]{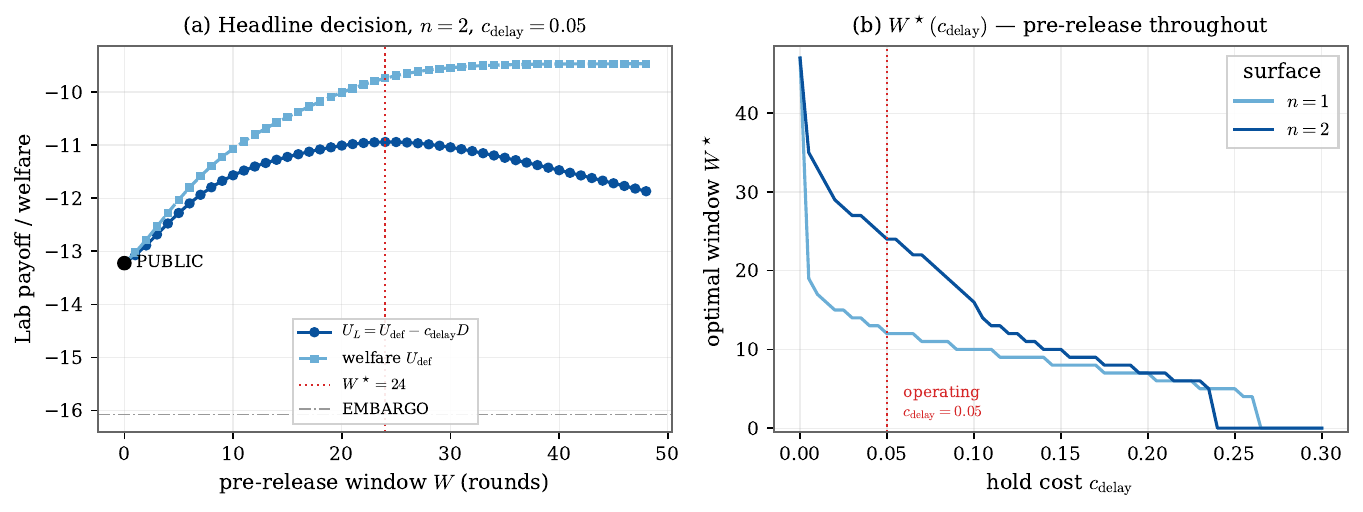}
\caption{Calibrated decision for Opus 4.6 $\to$ Mythos ($n=2$, $c_{\mathrm{delay}}=0.05$). (a) Lab payoff $U_{\Lab}$ and welfare $u_{\D}$ over the window $W$, peaking at the interior $W^\star{=}24$, above public release and embargo. (b) Optimal window $W^\star$ over the hold cost $c_{\mathrm{delay}}$, a pre-release beneficial for both $n{=}1$ and $n{=}2$.}
  \label{fig:release-decision}
\end{figure}

\vspace{-0.75em}
\paragraph{Robustness.}
The LLM-Delphi panel introduces epistemic uncertainty, reflected in the inter-expert standard deviations reported in \Cref{tab:delphi-capabilities}. We propagate this uncertainty by Monte Carlo to assess its effect on the optimal release strategy. For each model transition we draw the previous ($\bm{\kappa}^{\mathrm{prev}}$) and new ($\bm{\kappa}^{\mathrm{new}}$) capability vectors, sampling each component independently from a Gaussian with its elicited mean and standard deviation, truncated to $[0,1]$, and re-solve at $n=1$. A pre-release remains optimal at both $n=1$ and $n=2$ throughout the examined cost range (\Cref{fig:release-decision}b), although the window is longer at $n=2$, $24$ rather than $12$ rounds at the operating point, the value we report for the headline. Across all transitions a pre-release is the modal decision, selected in $80\%$ to $95\%$ of draws, most decisively for the largest offensive jump, Opus 4.6 $\to$ Mythos, at $95\%$, with a median window of $12$ rounds $([9,15])$ (\Cref{tab:robust}).

\begin{table}[t]
\centering
\caption{Release decision under calibrated uncertainty ($N=150$ draws per transition, $n=1$). $P(\mathsf{PRE})$ is the share of draws for which a pre-release is optimal. $W^\star$ is the median pre-release window and $\Delta u_\D$ the mean defender-welfare gain relative to public release, both reported with $[p_{10},p_{90}]$ intervals. Attacks averted is the mean reduction in attack frequency, relative to public release.}
\label{tab:robust}
\footnotesize
\setlength{\tabcolsep}{6pt}
\begin{tabular}{lcccc}
\toprule
Release transition & $P(\mathsf{PRE})$ & $W^\star\,[p_{10},p_{90}]$ & $\Delta u_\D\,[p_{10},p_{90}]$ & \makecell{Attacks\\averted} \\
\midrule
GPT-4o $\to$ o4-mini        & $0.873$ & $25\ [8,36]$  & $+4.06\ [0.15,8.43]$ & $41\%$ \\
o4-mini $\to$ Opus 4.5      & $0.853$ & $19\ [9,25]$  & $+2.94\ [0.00,5.56]$ & $32\%$ \\
Opus 4.5 $\to$ Opus 4.6     & $0.800$ & $11\ [2,15]$  & $+1.31\ [0.00,2.71]$ & $14\%$ \\
Opus 4.6 $\to$ Mythos Prev  & $0.953$ & $12\ [9,15]$  & $+2.17\ [0.66,3.66]$ & $23\%$ \\
\bottomrule
\vspace{-2em}
\end{tabular}
\end{table}

\vspace{-0.75em}
\paragraph{The release decision.}
Discovery $\kappa^{\mathrm{disc}}$ and exploit generation $\kappa^{\mathrm{egen}}$, the capabilities most influential for the adversary (\Cref{fig:cap-morris}), increase most sharply in the Opus 4.6 $\to$ Mythos Preview transition (\Cref{tab:delphi-capabilities}). Anthropic ran a pre-release program for this transition. Project Glasswing \cite{anthropic2026glasswing} gave selected partners early access to the then-unreleased Mythos Preview for vulnerability detection. Mozilla reportedly used that access to identify and remediate $271$ Firefox vulnerabilities before public release, roughly ten times the number found under Opus 4.6 \cite{holley_zero-days_2026}.
This transition is therefore a natural case to examine in detail. We solve it at the calibrated operating point of $n=2$ vulnerabilities (\Cref{subsec:parameterization}), where the adversary is fully two-route (\Cref{fig:reveng-regime}). Over $\mathcal{A}_{\Lab}$, the lab's Stackelberg solution is a pre-release window of $W^\star=24$ rounds. Under our calibration, this corresponds to approximately $168$ days of exclusive defender access. Relative to public release, it preserves $+3.48$ welfare and reduces attack frequency from $0.34$ to $0.27$ (\Cref{fig:release-decision}).

Our framework provides a structural interpretation of programs such as Glasswing. It identifies the capability gap as the welfare driver, predicts gains for offense-weighted releases, and converts the qualitative case for early access into a model-implied window. The round-to-wall-clock mapping is approximate, so the calibration identifies a welfare-beneficial strategy rather than a precise duration. The strategy is nevertheless robust, with a pre-release optimal in $80\%$ to $95\%$ of draws for every transition in \Cref{tab:robust}. With monetary values for per-round damage $d^{(v)}$ and delay cost $c_{\mathrm{delay}}$, which are treated here as a numeraire and a swept parameter, respectively, the model could express both the welfare gain and the optimal window in economic terms. 
\vspace{-0.75em}
\section{Discussion and Limitations}
\label{sec:discussion_limitations}
\vspace{-0.75em}
Our model provides a tractable framework for analyzing how frontier AI release policies shape defender welfare, but this tractability rests on several theoretical concessions and empirical approximations.
\vspace{-0.75em}
\paragraph{Game-theoretic abstractions.}
To guarantee a unique computable value, the inner game relies on three simplifications: full observability, a zero-sum payoff, and a finite horizon. The zero-sum reward assumes a pure damage-maximizing adversary and so establishes an upper bound on harm, since a real adversary bearing operational costs would attack less. Full observability lets the defender prioritize vulnerabilities by the adversary's progress and time patch shipping to limit exposure to reverse engineering, advantages a real defender would lack; this overstates the defender's strength but preserves the release decision, since protection derives from the capability gap rather than from information control. Finally, a finite horizon bounds the game to a single release cycle: each generation's capabilities are fixed per-round rates, and the policy sets only \emph{which} generation each side can access at each round, so AI advancement enters as the discrete $\bm{\kappa}^{\mathrm{prev}}\!\to\!\bm{\kappa}^{\mathrm{new}}$ jump.
\vspace{-0.75em}
\paragraph{Calibration.}
The delay and action costs cannot be grounded in public data, so we sweep them and report the window as a function of cost (\Cref{fig:release-decision}b) rather than a single value; the reported windows are therefore conditional on the chosen operating point, unlike calibrated parameters such as the adoption rate $\rho$. The capability rates $\bm{\kappa}$ are elicited by the LLM-Delphi procedure of Lorenz and Fritz \cite{lorenz_scalable_2026} rather than directly measured, following prior work that elicits risk estimates from expert panels \cite{barrett_toward_2025}; because the procedure maps benchmark pass rates to per-round rates, it could also accommodate evidence from scaffolded agentic systems, where frontier capability increasingly resides. The elicited values determine how large the protective gap is, not whether it protects; the latter holds across the elicited range (\Cref{subsec:prerelease}). Finally, the round-to-wall-clock mapping rests on sector-average remediation data and is unvalidated against any specific rollout, so the framework identifies an optimal release strategy, not a precise duration.
\vspace{-0.75em}
\paragraph{Future work.}
The model identifies the welfare-optimal pre-release window, but the lab weighs its private cost of delay against defender welfare, so the window it chooses can fall short of the social optimum. Aligning the two through mechanism design is the extension this work most directly motivates. Beyond it, the model can be extended toward real-world dynamics: heterogeneous defenders, richer release instruments such as security guardrails and performance throttling, repeated releases, and competition between labs.

\vspace{-0.75em}
\section{Conclusion}
\vspace{-0.75em}
Responsible disclosure has long given defenders a head start in the vulnerability lifecycle. Frontier AI erodes that advantage by granting both sides the same model, pushing the release decision upstream to the labs that build them. We model that decision as a bilevel three-player Stackelberg game whose leader determines the release strategy, setting both sides' capabilities and parameterizing a downstream zero-sum stochastic game between defender and adversary.

Defender welfare depends on the AI capability \emph{gap} between defender and adversary, not on the absolute capability level. Because a patch must be built and tested while an exploit need only be weaponized, the defender's extra pipeline steps turn equal AI capability gains into a smaller end-to-end speedup than the adversary's, so a timed pre-release of a frontier AI model to the defender creates a protective gap. Calibrated to successive frontier-model transitions via an LLM-Delphi elicitation method, a pre-release is the lab's equilibrium choice in $80$--$95\%$ of posterior draws.

For models that clear the deployment threshold but carry dual-use capability, the protective lever is sequencing access rather than withholding it entirely. Today's threshold-based safety frameworks do not capture this graduated control. Aligning the lab's private delay cost with the social optimum through mechanism design is the natural next step.

\vspace{-0.75em}
\begin{credits}
\subsubsection{\ackname} This work was partially funded by the European Union under Horizon Europe grants No.101214398 (ELLIOT) and No.101070617 (ELSA). Funded by the European Union. Views and opinions expressed are however those of the authors only and do not necessarily reflect those of the European Union or the European Commission. Neither the European Union nor the European Commission can be held responsible for them. It was also partially supported by the German Federal Ministry of Education and Research (BMBF) under grant AIgenCY (16KIS2012).
\end{credits}
%
%
%
\bibliographystyle{splncs04}
\bibliography{GameSec}
\end{document}